\documentclass[]{krakproc}
\usepackage{graphicx}
\begin{document}

\title{Overview of the ISM Phases: Evolution of Large and Small Scale Structures
in 3D high resolution HD and MHD simulations}
\author{Dieter Breitschwerdt$^1$ and Miguel A. de Avillez$^{1,2}$}
\institute{$^1$Institut f\"ur Astronomie, Universit\"at Wien,
     T\"urkenschanzstr.\ 17, A-1180 Wien, Austria \\
        email: breitschwerdt@astro.univie.ac.at\\
     $^2$Department of Mathematics, University of \'Evora,
              R. Rom\~ao Ramalho 59, 7000 \'Evora, Portugal \\
              email: mavillez@galaxy.lca.uevora.pt }
\markboth{D. Breitschwerdt \& M. Avillez}{Evolution of the ISM
\ldots}

\maketitle

\begin{abstract}
We review recent 3D high resolution numerical HD and MHD studies of
a multi-component and multi-phase ISM. The computational grid was
chosen large enough in the disk to represent typical ISM gas
patterns, and in the halo to include the full extent of the Galactic
fountain. The evolution of the ISM is driven by supernovae of all
types and followed for 400 Myrs, long enough to get rid of memory
effects of the initial distributions. The results are substantially
different from previous classical analytical models and entail: (i)
a very inhomogeneous distribution of ISM gas, (ii) a high level of
supernova driven turbulence, (iii) an active Galactic fountain flow,
reaching out to about 5 kpc on either side of the disk, (iv) a low
volume filling factor for the hot gas of about 20\%, (v) a very weak
correlation between magnetic field and gas density, $n$, for $n <
100 \, {\rm cm}^{-3}$, and, (vi) about 50\% of the gas disk mass
with $n < 100 \, {\rm cm}^{-3}$, residing in classically thermally
unstable temperature regimes.

\end{abstract}

\section{Introduction}

Modern history of the interstellar medium (ISM) commences with the
discovery of the hot component (HIM) with the Copernicus satellite,
revealing a diffuse wide-spread O{\sc vi} resonance line doublet in
absorption towards background stars (Jenkins \& Meloy 1974, York
1974). Earlier, following the discovery of the X-ray background by
Giacconi et al. (1962), diffuse soft X-ray emission below 2 keV was
observed (Bowyer et al. 1968). While much of its harder part has
been resolved into point sources according to deep field
observations with the Chandra (Mushotzky et al. 2000) and XMM-Newton
(Hasinger et al. 1998, 2001) X-ray observatories, the emission below
$\sim 1/2$ keV must be largely thermal in origin coming from a hot
plasma. It became therefore clear, even in the early days, that a
new ``phase'' of the ISM had been discovered, which should dominate
by volume, and due to the implied temperature range ($2 \times 10^5
\leq T \leq 2 \times 10^6$ K) must be mainly generated by shock
heating of supernova remnants (SNRs). We note parenthetically that
the gas causing the O{\sc vi} absorption cannot be the one emitting
in soft X-rays, since in collisional ionization equilibrium in the
latter O{\sc vii} would dominate O{\sc vi} by two orders of
magnitude.

It was shown by Cox \& Smith (1974) that a supernova (SN) rate of
$1$ per $50$ years is sufficient to maintain a large scale Galactic
tunnel network. A corollary of their work was that the H{\sc i}
should be distributed in filaments, some of them dense enough to be
associated with tunnel walls. The filamentary structure of the cold
gas is actually also found in our 3D high resolution simulations
(Avillez \& Breitschwerdt 2004, 2005), which will be discussed here
in some detail. The disk-halo circulation (or Galactic fountain,
cf.\ Shapiro \& Field 1976, Bregman 1980, Kahn 1981) reinforced by
the clustering of O and B stars in the disk, was certainly a salient
feature missing in the early studies. McKee \& Ostriker (MO, 1977)
have tried to put all the wisdom of the ISM into a coherent picture,
coined the three-phase ISM model. One of the most interesting
features is the transformation of the ``phases'' into one another by
different physical processes, such as e.g.\ radiative cooling, heat
conduction, photoionization and evaporation. In general ISM physics,
the term ``phases'' refers to the stable branches (upon entropy
perturbations) in the $p-T$-diagram. The underlying assumption of
the MO model is that the ISM is in global pressure equilibrium.

A heavily non-linear system like the ISM is susceptible to changes
in the initial conditions, as it is known from deterministic chaos
theory. Therefore the richness of phenomena cannot be covered in an
analytical model, and detailed numerical simulations are required.
It is therefore not surprising that the MO model fails, e.g., in the
predicted volume filling factors of the diffuse ionized gas (DIG)
and especially of the HIM; the latter is at least a factor of 2 too
high. This is partly due to the fact that clustered SNe were not
considered, as was done in the so-called ``chimney'' model (Norman
\& Ikeuchi 1989). Here break-out and establishment of the fountain
flow was accounted for, but the physics of break-out was rather
idealistic.

Not surprisingly, all analytic models and low resolution 2D
simulations have underestimated the r\^ole of turbulence and have
missed the amount of gas which resides in thermally unstable
regimes, as will be discussed in some detail in this review.

\section{3D High Resolution Simulations}
\label{hrs}
\subsection{The model}
\begin{table}[thbp]
\centering \caption{Average volume filling factors of the different
ISM phases for different SN rates. The average was calculated using
101 snapshots (of the 1.25 resolution runs) between 300 and 400 Myr
of system evolution with a time interval of 1 Myr. \label{Avervff} }

\begin{tabular}{ccccc}
\hline \hline $\sigma$$^a$ & $\langle f_{v, cold}\rangle$$^b$ &
$\langle f_{v, cool}\rangle$$^c$ &  $\langle f_{v, warm}\rangle$$^d$ & $\langle f_{v, hot}\rangle$$^e$ \\
\hline
1 & 0.171 & 0.354 & 0.298 & 0.178 \\
2 & 0.108 & 0.342 & 0.328 & 0.223 \\
4 & 0.044 & 0.302 & 0.381 & 0.275 \\
8 & 0.005 & 0.115 & 0.526 & 0.354 \\
16 & 0.000 & 0.015 & 0.549 & 0.436 \\
\hline
\multicolumn{5}{l}{$^a$ SN rate in units of the Galactic SN rate.}\\
\multicolumn{5}{l}{$^b$ $T<10^{3}$ K.}\\
\multicolumn{5}{l}{$^c$ $10^{3} <T\leq 10^{4}$ K.}\\
\multicolumn{5}{l}{$^d$ $10^{4} <T\leq 10^{5.5}$ K.}\\
\multicolumn{5}{l}{$^e$ $T> 10^{5.5}$ K.}\\
\end{tabular}
\end{table}
The hydrodynamical (HD) and magnetohydrodynamical (MHD) equations
are solved on a Cartesian grid of $0\leq (x,y)\leq 1$ kpc size in
the Galactic plane and extending $-10\leq z \leq 10$ kpc into the
halo (for details see Avillez 2000, and Avillez \& Breitschwerdt
2004). The kernel of the model consists of a three-dimensional HD or
MHD code that uses adaptive mesh refinement (AMR) in a block-based
structure in combination with Message Passing Interface (MPI)
suitable for massive parallel computations. This will ensure the
full capturing of the disk-halo-disk cycle, i.e.\ the Galactic
fountain flow. The grid is centered on the solar circle, with a
finest adaptive mesh refinement resolution of 0.625 (HD) and 1.25 pc
(MHD), respectively, in the layer $-1\leq z\leq 1$ kpc. Following
the observations, basic physical ingredients of the model are: (i)
massive star formation in regions of converging flows ($\nabla\cdot
\vec{v}<0$, $\vec{v}$ being the gas velocity) where density and
temperature are $n\geq 10$ cm$^{-3}$ and T$\leq 100$~K,
respectively; (ii) the distribution of masses and individual stellar
life times are derived from a Galactic initial mass function for
massive stars; (iii) SN explosions occur at a Galactic rate and with
a canonical energy of $10^{51}$ erg (including the scale height
distributions of types Ia, Ib and II); (iv) the gas is immersed in a
gravitational field provided by the stellar disk, following Kuijken
\& Gilmore (1989); (v) radiative cooling (assuming an optically thin
gas in collisional ionization equilibrium) with a temperature
cut-off at 10 K, and uniform heating due to starlight varying with
$z$ (Wolfire et al. 1995); (vi) in case of the MHD simulation, an
initially disk parallel magnetic field composed of random ($B_r$)
and uniform ($B_u$) components, with total field strength of
$4.5~\mu$G ($B_u=3.1$ and $B_r=3.3~\mu$G). The establishment of the
fountain flow up to vertical heights of $5 - 10$ kpc takes about
$100 - 200$ Myr, so that a simulation time of 400 Myr seems adequate
in order to recover the main features. Boundary conditions are
periodic on the side walls of the computational box and outflow on
its top and bottom.

\subsection{The results}

\begin{figure*}[!ht]
  \centering
  \includegraphics[width=0.3\hsize,angle=-90]{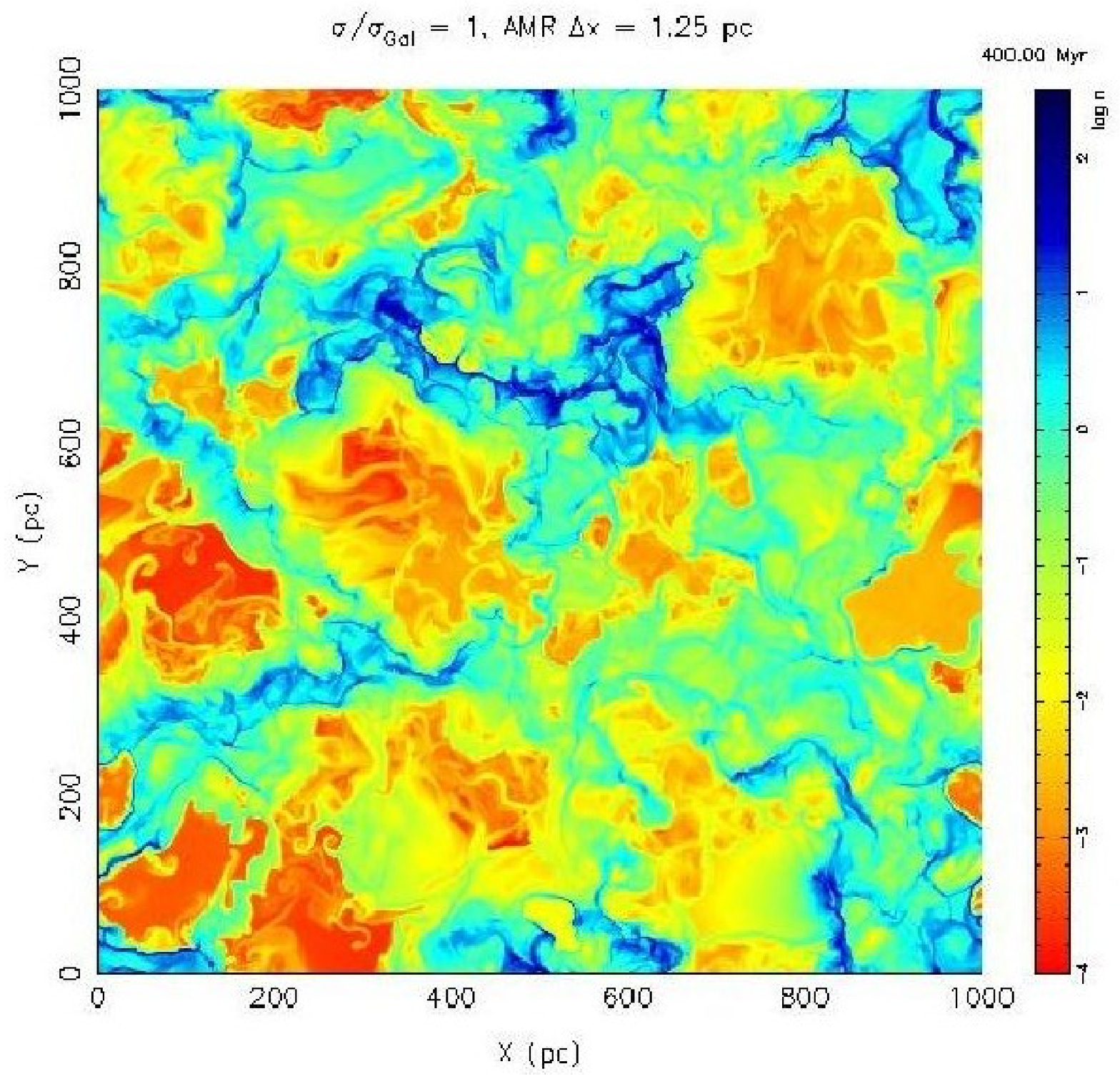}%
  \includegraphics[width=0.3\hsize,angle=-90]{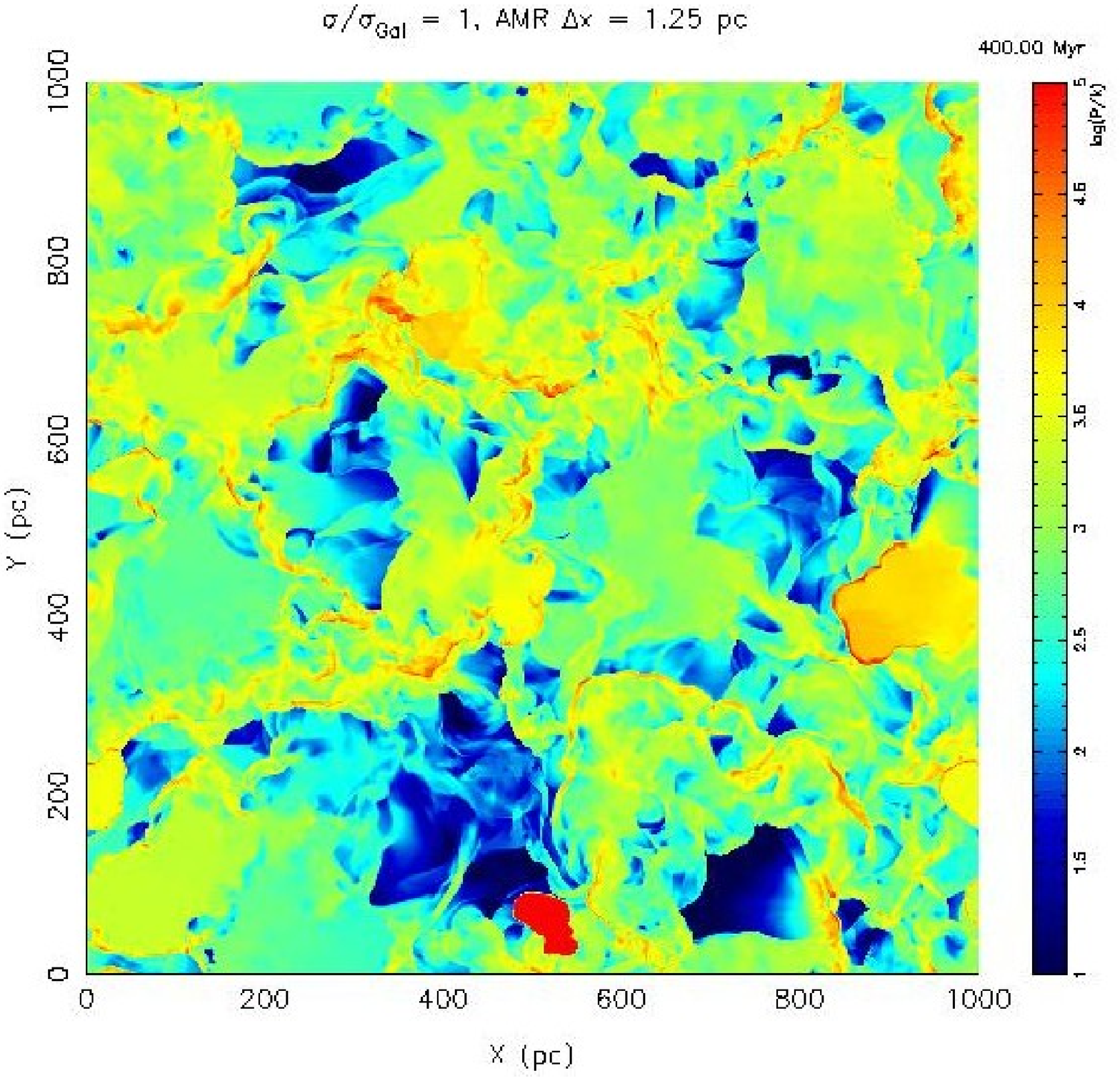}%
  \includegraphics[width=0.3\hsize,angle=-90]{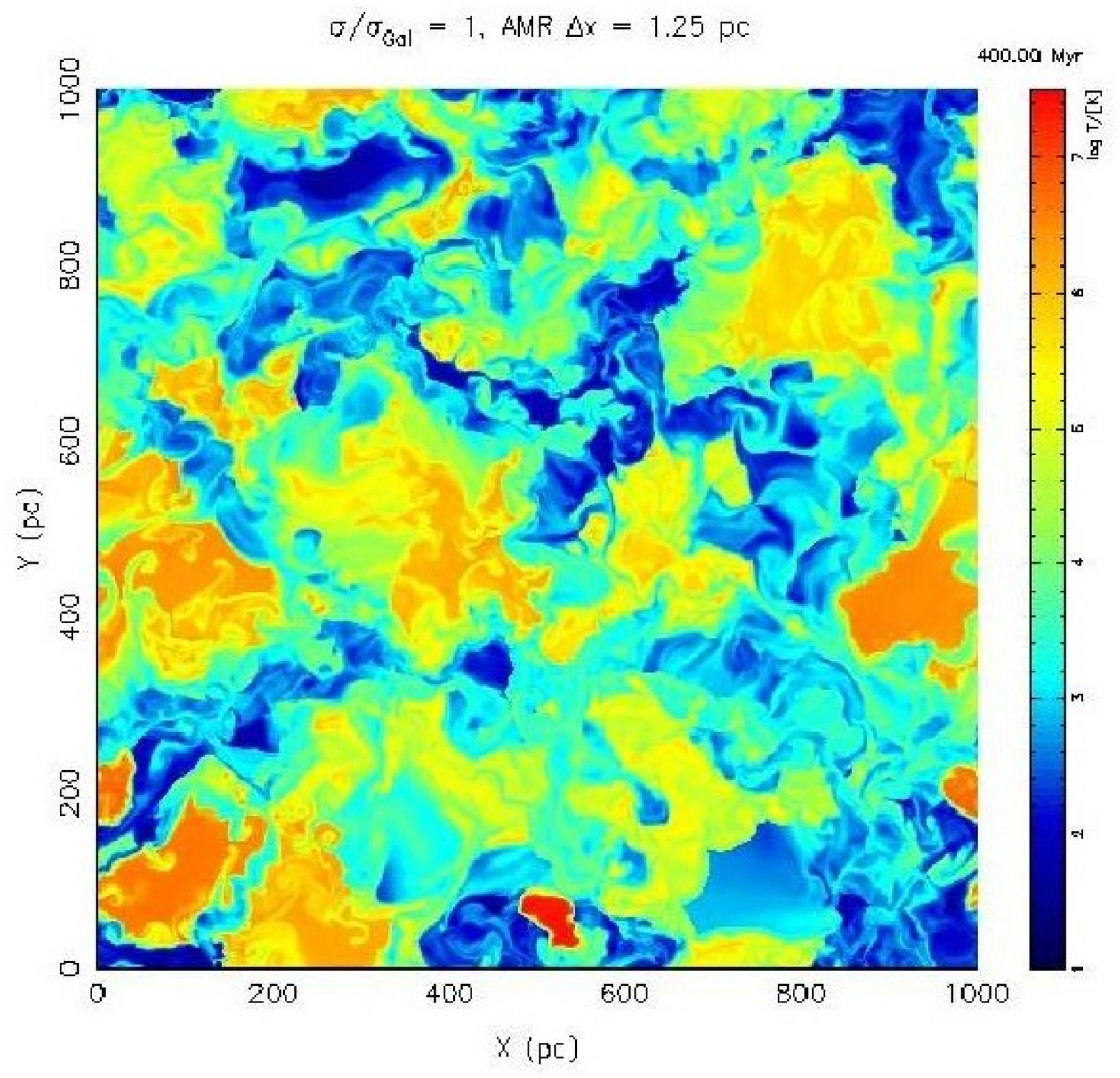}%


\vspace*{0.25cm}
  \includegraphics[width=0.3\hsize,angle=-90]{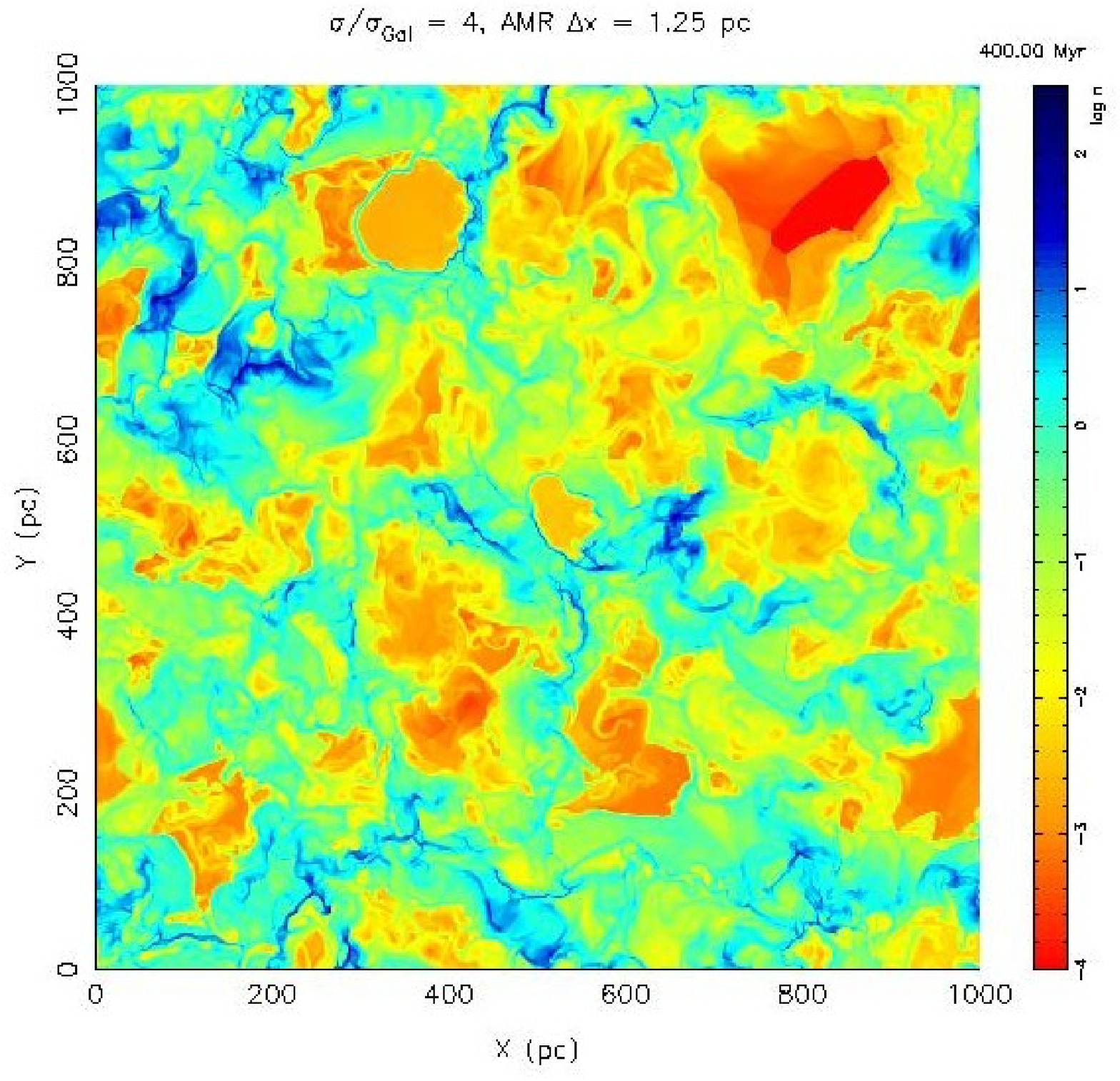}
  \includegraphics[width=0.3\hsize,angle=-90]{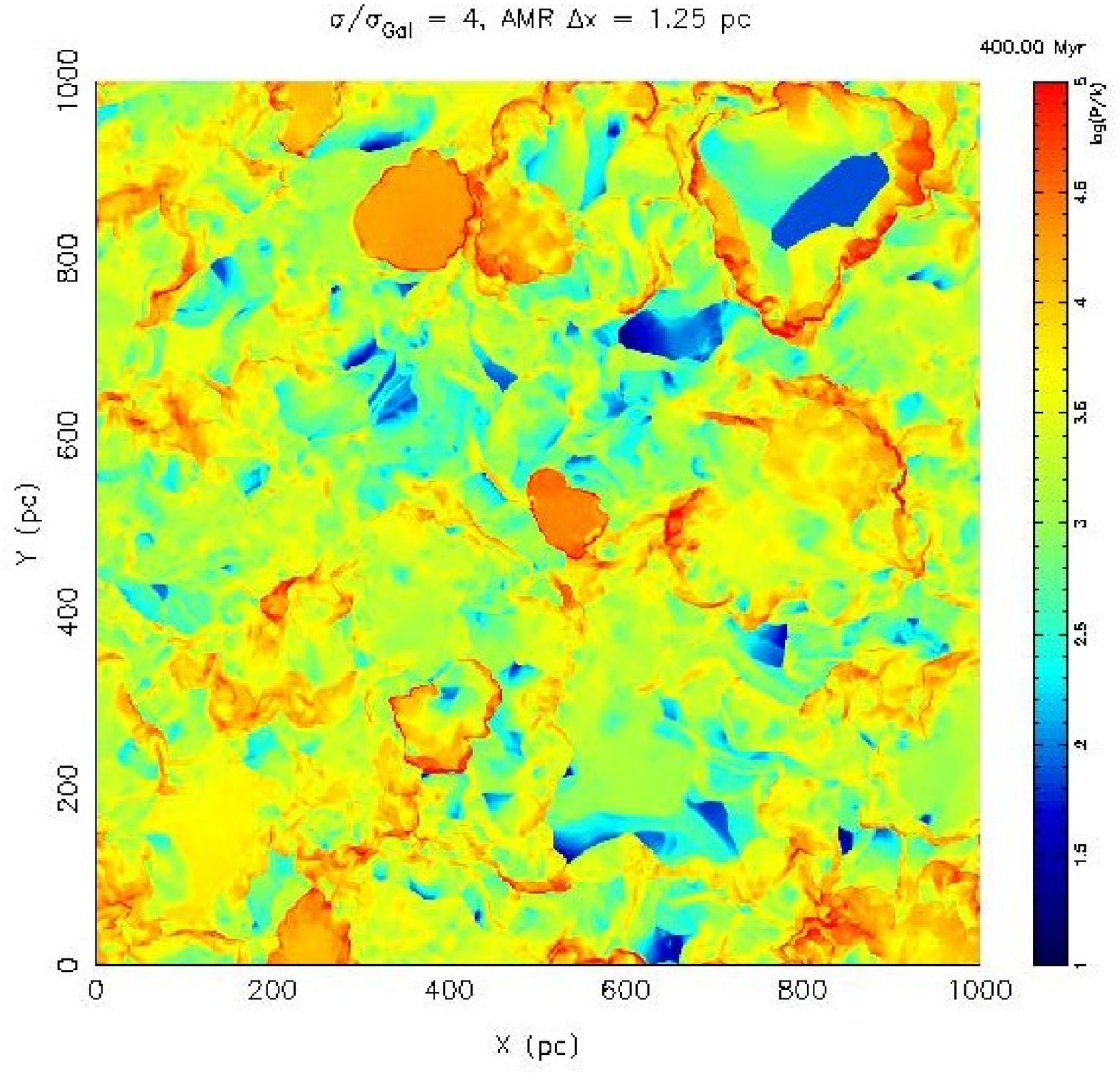}
  \includegraphics[width=0.3\hsize,angle=-90]{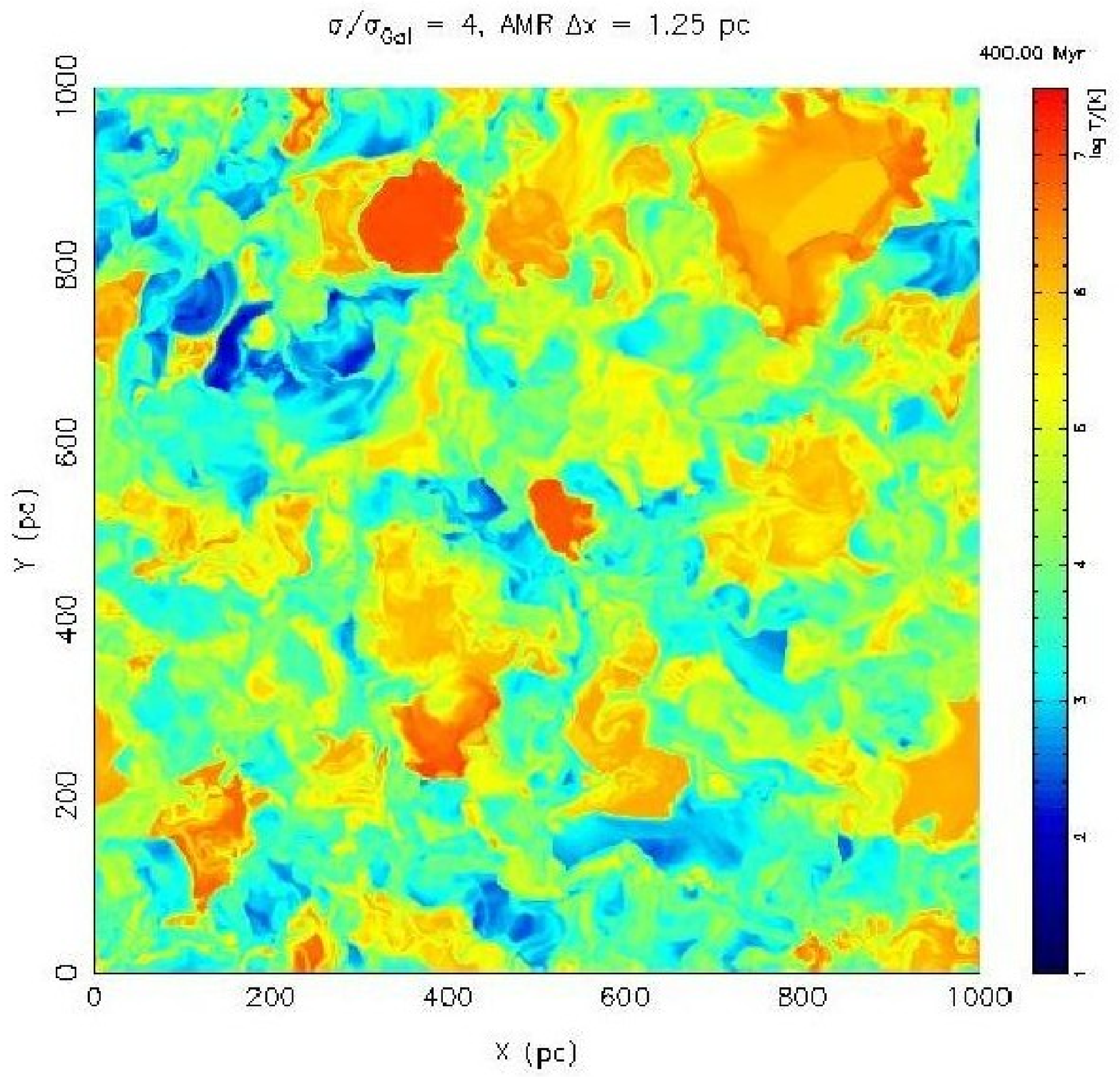}

\caption{Two dimensional cuts, through the 3D data cube, showing $n$
(density, \textbf{left} column), $P/k$ (pressure, \textbf{middle}
column) and $T$ (temperature, \textbf{right} column) distribution in
the Galactic plane for supernova rates $\sigma/\sigma_{Gal}=1$ (top
row) and 4 (bottom row). The colour coding of the logarithmic scale
refers to red for low density/high pressure/high temperature and
dark blue for high density/low pressure/low temperature,
respectively.
\label{1.25pc-GP}}
\end{figure*}
In comparison to other authors, our results cover a large fraction
of the galactic disk and the halo and yet retain a high spatial
resolution, and follow the evolution over a time scale long enough
to have erased all memory effects imprinted by the initial
conditions. These have always to be chosen in an artificial way,
since we are only capturing a small window of a galaxy's long
history. In Fig.~\ref{1.25pc-GP} the distribution of the major
quantities (for HD simulations), characterizing the physical state
of the ISM in the disk, is shown: gas density $n$, pressure $P/k$
(where $k$ is Boltzmann's constant), and the gas temperature $T$.
The first row represents these variables for the Galactic SN rate
$\sigma_{gal}$ and the second for $\sigma = 4 \sigma_{gal}$.
\begin{figure*}[!t]
\centering
\includegraphics[width=1.0\hsize,angle=-90]{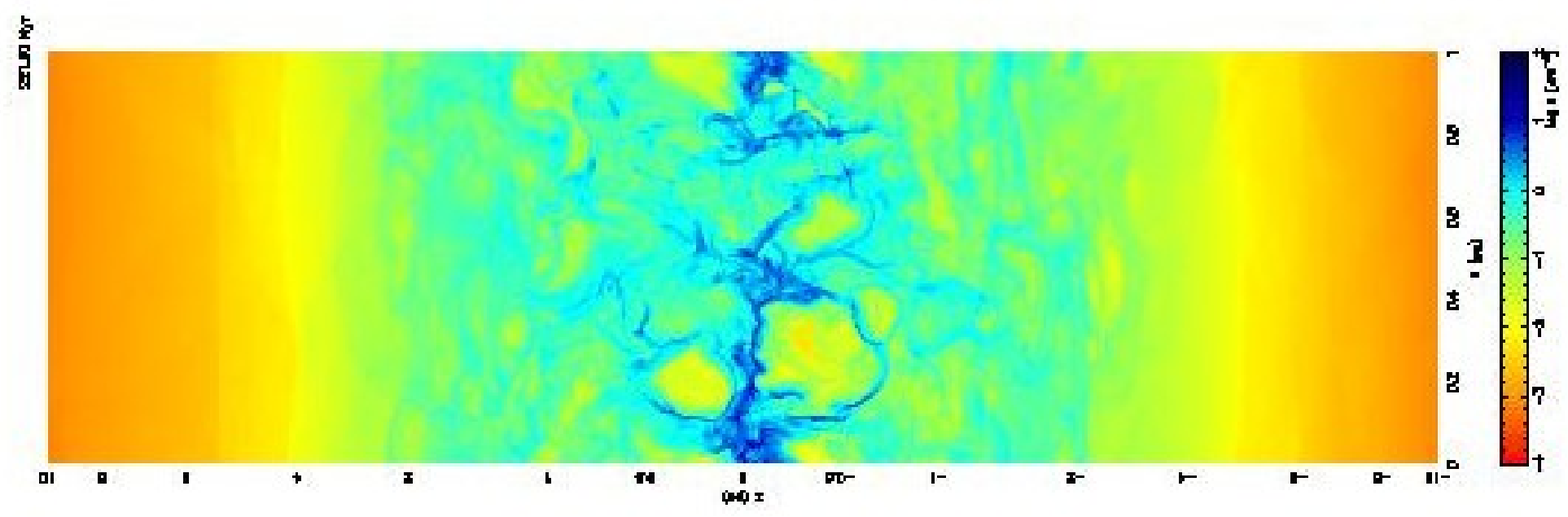}
\includegraphics[width=1.0\hsize,angle=-90]{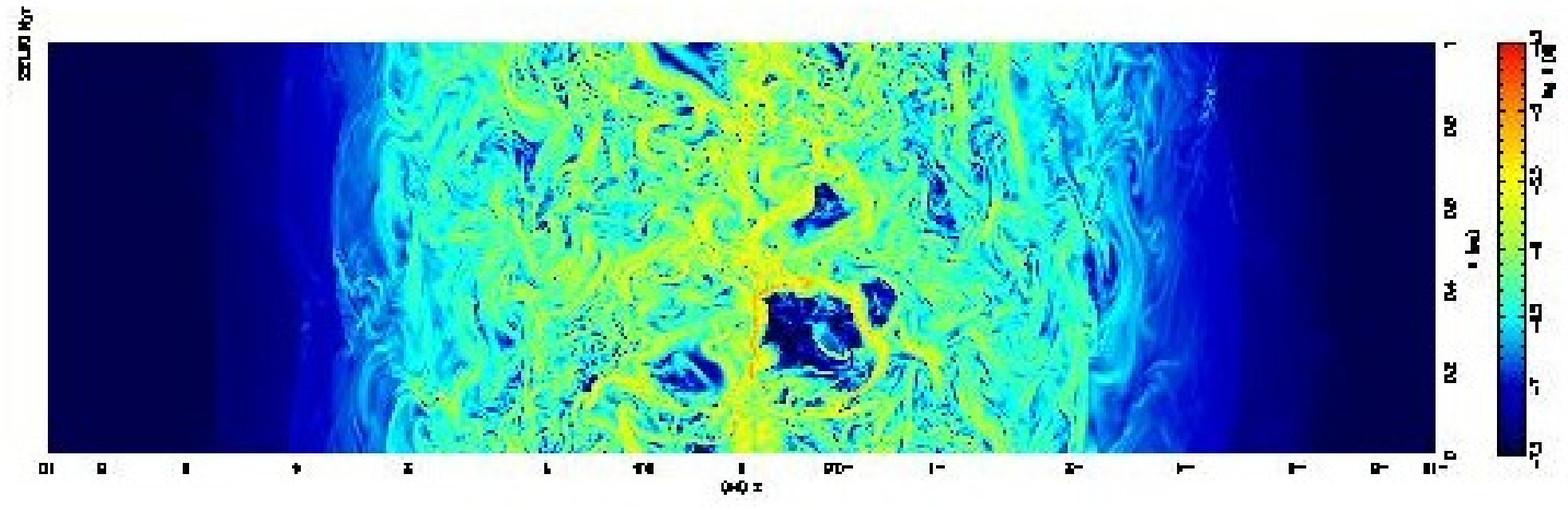}
\caption{Slices through the 3D data cube showing the vertical
(perpendicular to the midplane) distribution of the density
(\textbf{left}) and magnetic field (\textbf{right}) at time $t=330$
Myr. Colour coding refers to red as low density/high field, and dark
blue to high density/low field. \label{dh} }
\end{figure*}
Looking at the density and temperature, we see that the structure of
the ISM is very \emph{inhomogeneous}. There are regions of SNR and
superbubble (SB) size at very low density and high temperature,
interspersed with shock compressed layers, originating from
colliding laminar flows. The pressure $P/k \sim 3000 \, {\rm K}\,
{\rm cm}^{-3}$ is considerably lower than in most (analytical)
models and it is not at all uniform, showing substantial variation
of more than one order of magnitude; note that this was one of the
basic assumption of the three-phase MO model. Another most
remarkable result is that almost half of the ISM mass resides in
thermally unstable phases. How is this possible? The classical Field
(1965) criterion, according to which instability is expected to set
in, if $\left(\frac{\partial \mathcal{L}}{\partial T} \right)_P <0$,
where $\mathcal{L}$ is the heat loss function per unit mass, is
insufficient, since a gas element does not just sit there and cool
down. On the contrary, due to the high level of turbulence, it is
constantly shuffled around and finds itself in widely different
environments. Thus, one possibility to get around catastrophic
energy losses is if the typical eddy crossing time is less than the
radiative cooling time, i.e., $\tau_{\rm eddy} \sim \lambda/\Delta u
\ll \tau_{\rm cool}$ (with $\Delta u$ being the turbulent velocity
fluctuation amplitude and $\lambda$ the eddy scale). For Kolmogoroff
turbulence, which seems to exist in a variety of different ISM
environments, we obtain in the inertial range for wavelengths
$\lambda$, $\lambda \sim 10^{18}- 10^{19}$ cm, for values
corresponding to the warm neutral medium (WNM) of about $\sim 1000$
K (cf.\ Avillez \& Breitschwerdt 2005). We find that $49\%$ of the
mass in the disk (see Table~\ref{table_mhd}) is concentrated in the
classical thermally \emph{unstable} regime $200< T \leq 10^{3.9}$ K
with $\sim 65\%$ of the warm neutral medium (WNM) mass enclosed in
the $500\leq T \leq 5000$ K gas, consistent with recent observations
by Heiles \& Troland (2003).

From Table~\ref{Avervff} the interesting result can be deduced, that
even for SN rates as high as 16 times the Galactic rate
(corresponding already to starburst conditions), the volume filling
factor of the hot gas remains below 50\%. In our Galaxy its values
is only 17\% in the HD and 21\% in the MHD simulations
(cf.~Table~\ref{table_mhd}), and therefore considerably lower than
the 50-70\% in the MO model. The main reason for this discrepancy
lies in the fact, that clustered SN explosions can enforce break-out
of SBs and thus establish a fairly efficient Galactic fountain.

The MHD simulations have been carried out under the same general
conditions as the HD calculations, except that the resolution limit
was 1.25 pc. After several stellar cycles the disk gets disrupted
due to SNR and SB break-out. The left panel of Fig.~\ref{dh} shows
the presence of a wiggly and thin disk of cold gas overlayed by a
frothy thick disk composed of neutral (light blue) and ionized
(greenish) gas. The break-out of SBs can be nicely seen, especially
in the southern hemisphere, where large bubbles of diameters of
$\sim 500$ pc expand into the halo, one even opening up. The
magnetic field map shows the thin magnetized disk overlayed by
beautiful magnetic loops, seen in Fig.~\ref{dh} (right) in
projection stacked on top of each other, magnetic islands, and
clouds wrapped in field lines moving downwards. There is also cold
gas descending along the loops. From these maps it can be
immediately seen that the simulations must take into account very
large heights on either side of the midplane, allowing for the setup
of the disk-halo-disk circulation.
Considering the controversial discussions during the late 80's,
mainly based on analytical calculations and 2D simulations, whether
SBs can break out of the disk, and finally blow out into the halo,
this problem seems to be resolved now. Unless a galaxy has an
extremely strong magnetic field, and if the field strength initially
decreases with $z$ proportional to the gas density, the set-up of
the galactic fountain flow can be at most delayed but not inhibited,
as can be seen from Fig.~\ref{dh}.
\begin{figure*}[thbp]
\centering
\includegraphics[width=0.49\hsize,angle=-90]{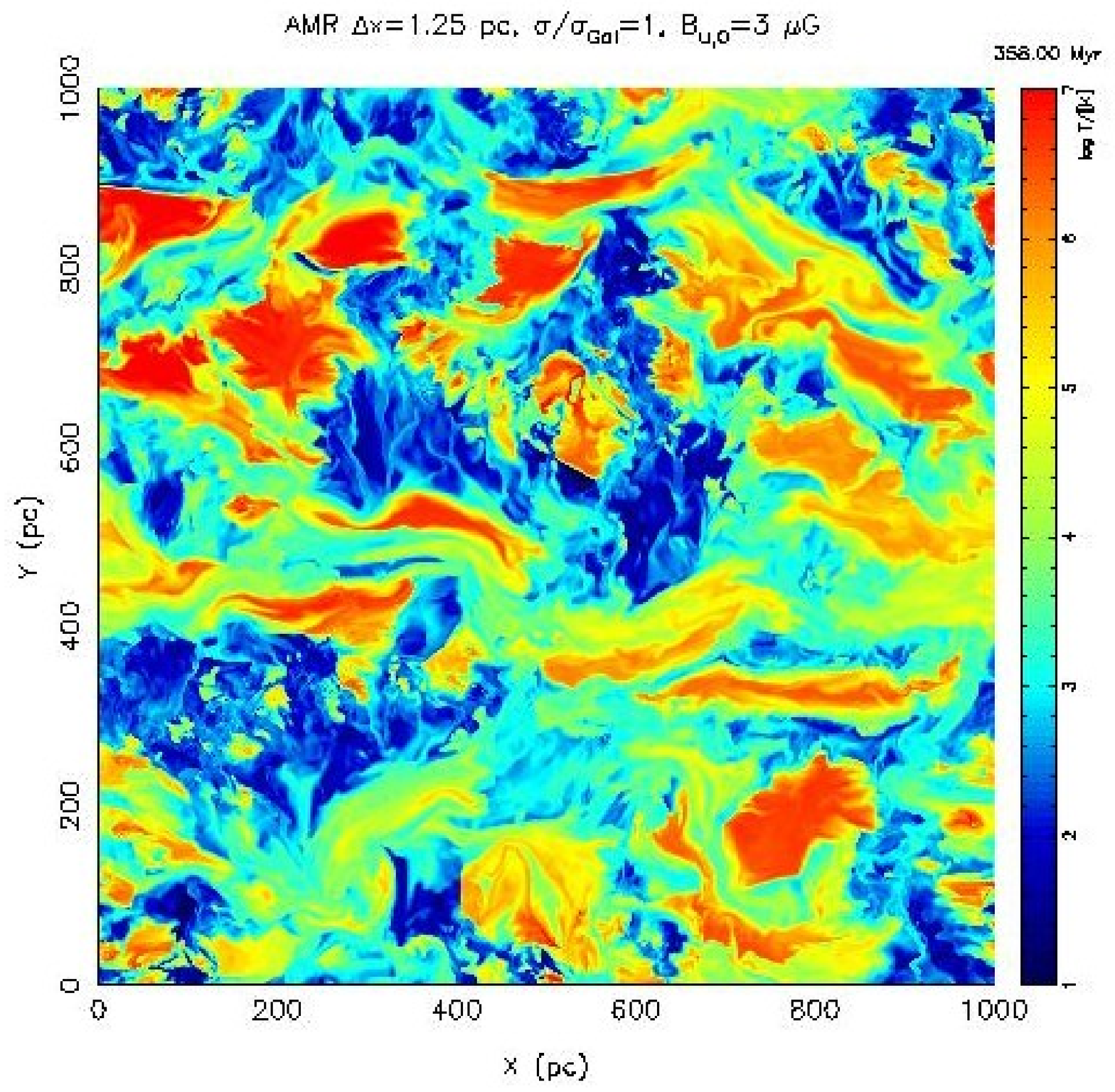}%
\includegraphics[width=0.49\hsize,angle=-90]{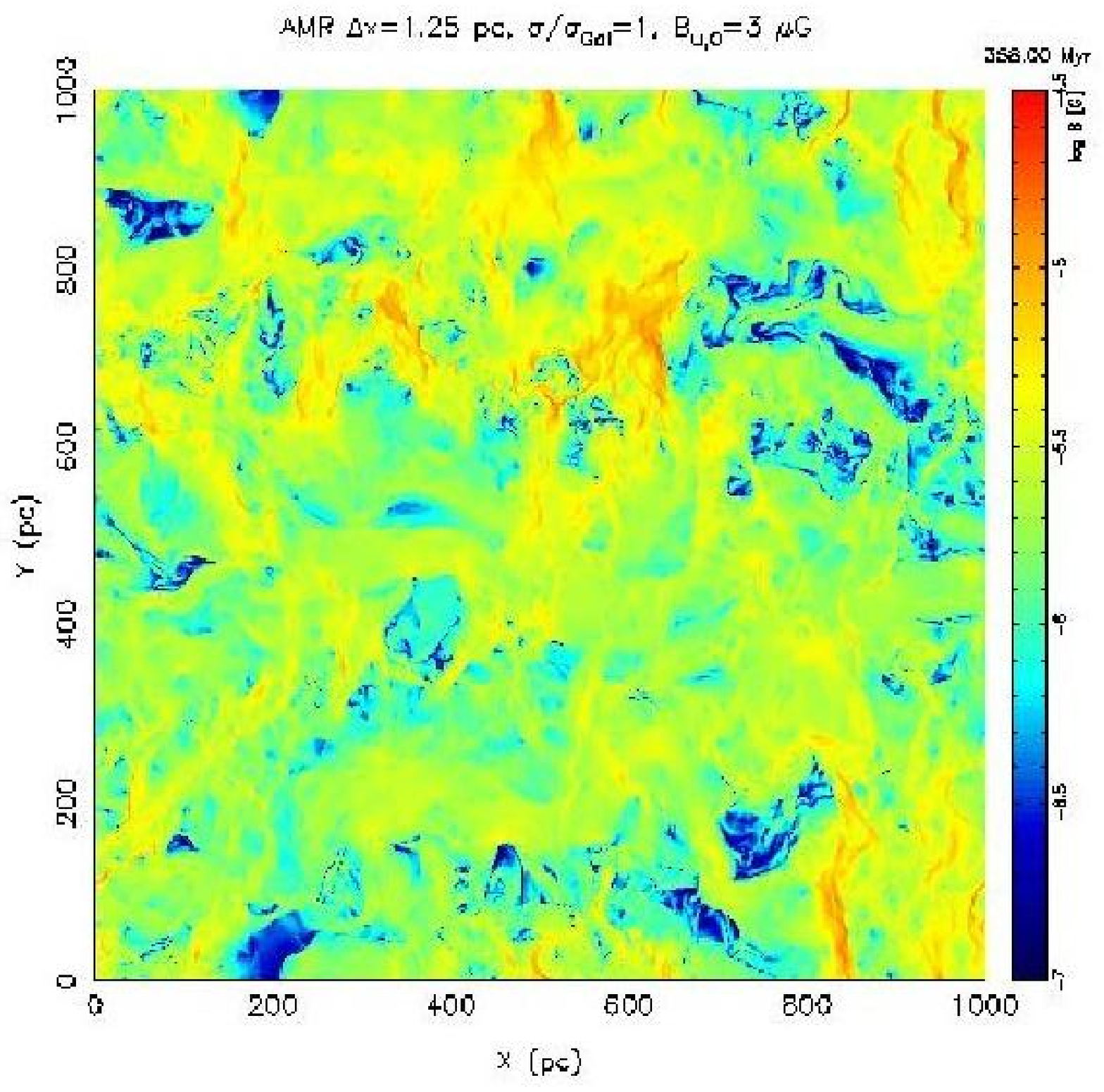}
\caption{2D slices through the 3D data cube at $z=0$ (Galactic
midplane), showing temperature (\textbf{left}) and magnetic field
(\textbf{right}) after 358 Myr of MHD evolution. Colour bars
indicate the logarithmic scale of each quantity. Red refers to
highest temperature/highest field, while dark blue refers to the
lowest temperature/lowest field strength. \label{fig1} }
\end{figure*}

\begin{table}[!hbtp]
\centering \caption{ Summary of average values of volume filling
factors, mass fractions and root mean square velocities of the disk
gas at the different thermal regimes for HD and MHD
runs\label{table_mhd}}
\begin{tabular}{c|cc|cc|cc}
\hline \hline
T  & \multicolumn{2}{c|}{$\langle \mbox{f}_{\mbox{v}}\rangle $$^{a}$ [\%]}& \multicolumn{2}{c|}{$\langle \mbox{f}_{\mbox{M}}\rangle $$^{b}$ [\%]} & \multicolumn{2}{c}{$\langle \mbox{v}_{\mbox{rms}}\rangle$$^{c}$}\\
\cline{2-7}
[K] & HD & MHD & HD & MHD & HD & MHD\\
\hline
$<200$ K              & ~5 & ~6 &  44.2 & 39.9 & ~7 & 10 \\
$200-10^{3.9}$        & 46 & 29 &  49.0 & 43.7 & 15 & 15 \\
$10^{3.9}-10^{4.2}$   & 10 & 11 &  ~4.4 & ~8.5 & 25 & 21 \\
$10^{4.2}-10^{5.5}$   & 22 & 33 &  ~2.0 & ~7.4 & 39 & 28 \\
$>10^{5.5}$           & 17 & 21 &  ~0.3 & ~0.5 & 70 & 55 \\
\hline
\multicolumn{7}{l}{$^a$ Occupation fraction.}\\
\multicolumn{7}{l}{$^b$ Mass fraction.}\\
\multicolumn{7}{l}{$^c$ Root mean square velocity in units of km
s$^{-1}$.}
\end{tabular}
\end{table}
\begin{figure}[thbp]
\centering
\includegraphics[width=0.49\hsize,angle=0]{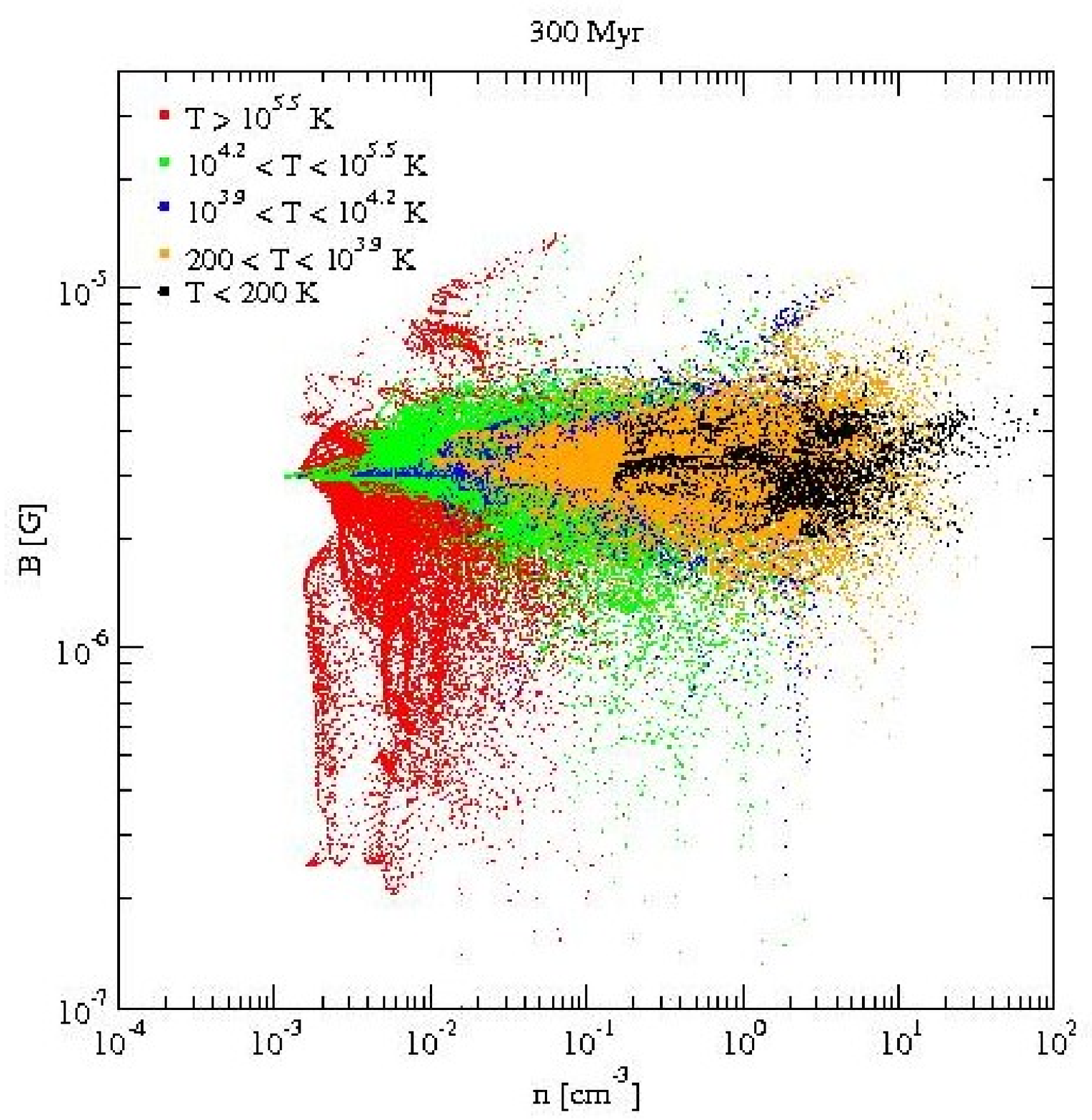}
\includegraphics[width=0.49\hsize,angle=0]{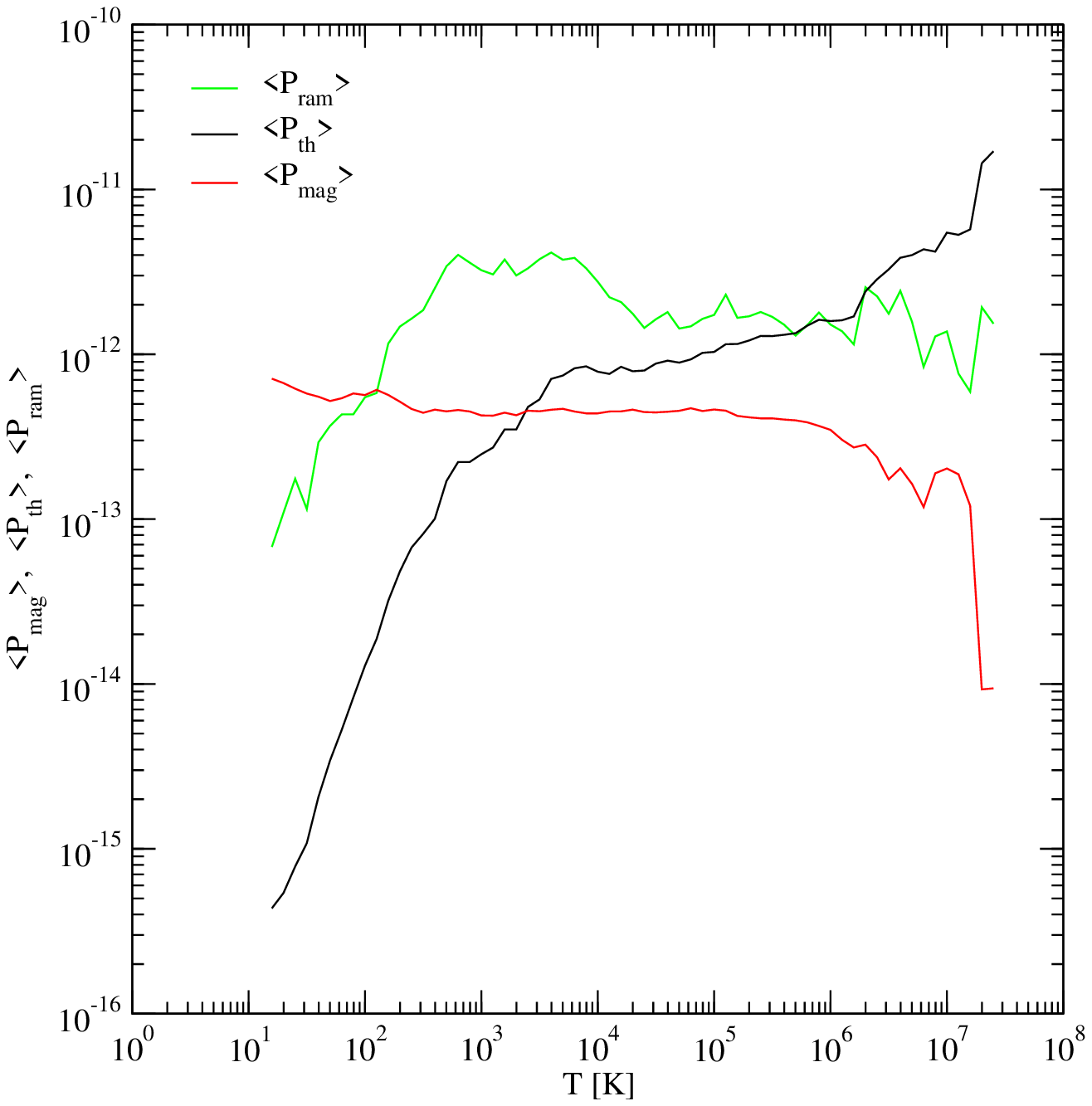}
\caption{\textbf{Left:} Scatter plot of B versus $n$ for $T \leq
10^{2}$
  (black), $10^{2}<T \leq 10^{3.9}$ (yellow), $10^{3.9}<T \leq
  10^{4.2}$ (blue), $10^{4.2}<T \leq10^{5.5}$ K (green), and
  $T >10^{5.5}$ K (red) regimes at 400 Myr of disk evolution. The
  points in the plot are sampled at intervals of four points in each
  direction.
  \textbf{Right:} Comparison of the average pressures $\langle$P$_{ram}\rangle$
(green), $\langle$P$_{th}\rangle$ (black), and
$\langle$P$_{mag}\rangle$ (red) as function of the temperature (in
the simulated disk $\left|z\right|\leq 250$ pc) averaged over
temperature bins of $\Delta \log T =0.1$ K.
  \label{scatterDB} }
\end{figure}
In the disk itself, the distribution of hot gas shows some features
of banana-shaped SBs (see Fig.~\ref{fig1} left), as would be
expected for an evolved bubble, when magnetic tension overcomes
internal pressure, and the bubbles can indeed shrink in the
direction perpendicular to the field wrapped around the shell (see
also Ferri\'{e}re et al. 1991). Although there is some general field
alignment of structures, the field itself is fairly non-uniform,
exhibiting small-scale structures (see Fig.~\ref{fig1} right).

As the magnetic lines of force are frozen into the ISM plasma, there
are some general arguments, that there should be a correlation, $B
\propto \rho^{\alpha}$, between mass density $\rho$ and magnetic
Field strength $B$. Chandrasekhar and Fermi (1953) have argued that
if the field is coupled to randomly moving gas clouds, the Alfv\'en
speed will be proportional to the turbulent velocity, and hence $B
\propto \rho^{1/2}$. Extracting a scatter plot of $B$ versus $n$
from our data cube, it can be seen that during the evolution of the
system the field strength broadened its distribution, spanning two
orders of magnitude after 400 Myr. Dense regions have stronger field
strengths and rarefied regions weaker strengths.
Fig.~\ref{scatterDB} (left) shows almost no correlation between
magnetic field strength and gas density except to some extent for
the cold dense regions (see also Heitsch et al. 2004).

We may finally ask ourselves what is actually driving the ISM flows,
and also the turbulence. In Fig.~\ref{scatterDB} (right) we have
plotted the volume averaged pressures (magnetic pressure in red,
thermal pressure in black, and ram pressure in green) against
temperature, in order to get an idea of the driving forces for the
various ISM phases and the unstable intermediate regimes. The result
can be summarized as follows: for low temperatures, magnetic
pressure dominates, whereas for high temperatures, clearly thermal
pressure is the main driving agent. But for all temperatures between
100 K and $10^6$ K ram pressure dominates.

\section{Conclusions}
\label{conc}
The ISM is a complex and highly non-linear system of gas, dust,
magnetic field and cosmic rays, powered by stellar photons, winds,
and most importantly, by supernova explosions. It is therefore not
surprising, that it shows a large variety of phenomena, e.g.\ phase
transitions, filamentary structure of cold and warm gas, highly
turbulent flows etc.. From analytical calculations, one can derive
the existence of phases, and their respective volume filling
factors, under certain assumptions, such as global pressure
equilibrium. However, quantitative estimates are rather difficult to
obtain, and can even be misleading, because some of the assumptions
are questionable. We have therefore followed a new approach, where
the evolution of a representative section of the magnetized gas disk
and its overlying halo are followed for an evolution time of 400
Myrs, with high spatial subparsec resolution. We thus find a highly
turbulent medium, containing about half of its mass in thermally
unstable temperature regimes, and exhibiting a large scale Galactic
fountain. We believe that the basic features of the ISM are captured
in our simulations. Future work includes the effects of e.g.\ cosmic
rays, non-equilibrium cooling and self-gravity of the gas on the
local and global ISM.

\textbf{Acknowledgements.} DB thanks the organizers for hospitality
and financial support.

\end{document}